\documentclass{webofc}
\usepackage[varg]{txfonts}   % Web of Conferences font
\usepackage{graphicx,epsfig,color}
\newcommand{\bc}{\begin{center}}
\newcommand{\ec}{\end{center}}
\newcommand{\be}{\begin{equation}}
\newcommand{\ee}{\end{equation}}
\newcommand{\bea}{\begin{eqnarray}}
\newcommand{\eea}{\end{eqnarray}}
\newcommand{\ba}{\begin{array}}
\newcommand{\ea}{\end{array}}
\newcommand{\lb}{\label}
\newcommand{\rf}{\ref}
\newcommand{\bfg}{\begin{figure}[htbp]}
\newcommand{\efg}{\end{figure}}
\begin{document}
\title{Impact of clustering inside compact tetraquarks}

\author{\firstname{Hagop} \lastname{Sazdjian}\inst{1}
\fnsep\thanks{\email{sazdjian@ijclab.in2p3.fr}}
}

\institute{Universit\'e Paris-Saclay, CNRS/IN2P3, IJCLab, 91405
Orsay, France}

\abstract{%
Due to the reducibility of tetraquark operators into mesonic
clusters, a strong interplay exists in tetraquarks between 
compact and molecular structures. This issue is studied within
an effective field theory approach, where the compact tetraquark
is treated as an elementary particle. Under the influence of the
coupling to the mesonic clusters, an initially formed compact
tetraquark bound state is deformed towards a new structure
of the molecular type, having the attributes of a shallow bound
state.
}
\maketitle
\section{Introduction} \lb{s1}

There has been growing experimental evidence, during the last two
decades, about the existence of exotic hadrons, also called 
multiquark states, containing more valence quarks than ordinary
mesons ($\overline qq$) and baryons ($qqq$).
Prototypes are tetraquarks with valence quark structure
$\bar q\bar qqq$, pentaquarks, with structure $\overline qqqqq$,
hexaquarks, with structure $qqqqqq$ or $\bar q\bar q\bar qqqq$
\cite{Choi:2003ue,Aubert:2003fg,Besson:2003cp,Aubert:2005rm,
Ablikim:2013mio,Liu:2013dau,Ablikim:2013wzq,Aaij:2014jqa,Aaij:2015tga,
Aaij:2020fnh,Aaij:2020ypa,Wu:2020hmk,SpadaroNorella:2021yje,
LHCb:2021auc,Gershon:2022xnn}.
\par
However, contrary to ordinary hadrons, multiquark states are not
color-irreducible, in the sense that they can be decomposed along
a finite number of combinations of ordinary mesonic or baryonic
clusters \cite{Jaffe:2008zz,Lucha:2019cdc}.
Schematically, ignoring here flavor and spin indices, a tetraquark
interpolating operator, $(\bar q\bar qqq)$, which is globally color
invariant, can be decomposed by means of Fierz rearrangements
into a form where clusters of mesonic operators have emerged:  
\be \lb{e1}
(\bar q\bar qqq)=\sum(\bar qq)(\bar qq),
\ee
where the parentheses signify color invariance of the included
operator.
Similar decompositions can be done with the pentaquark and hexaquark
interpolating operators:
\bea
\lb{e2}
& &(\overline qqqqq)=\sum(\overline qq)(qqq),\\
\lb{e3}
& &(qqqqqq)=\sum(qqq)(qqq),\ \ \ \ \  
(\bar q\bar q\bar qqqq)=\sum(\bar q\bar q\bar q)(qqq)
+\sum(\bar qq)(\bar qq)(\bar qq).
\eea
\par
Hadronic clusters, being color-singlets, mutually interact by means
of short-range forces, like meson-exchanges or contacts. They would
then form loosely bound states, in similarity with atomic
molecules. These are called \textit{hadronic molecules} or
\textit{molecular states} \cite{Voloshin:1976ap,
DeRujula:1976zlg,Tornqvist:1993ng}. 
\par
In contrast, multiquark states, formed directly from confining
interactions acting on all quarks, would form \textit{compact bound
states}, called \textit{compact multiquark states}
\cite{Jaffe:2003sg,Shuryak:2003zi,Maiani:2004vq,Maiani:2005pe}.
\par
Multiquark states can thus be formed by two different mechanisms,
each leading to a different structure. The issue is to find, by
theoretical justification, also guided by experimental data,
the most faithful description.
\par
In the following, we concentrate on tetraquark states, ignoring
details coming from quark flavors and spins, which do not play
a fundamental role.
\par

\section{Energy balance} \lb{s2}

A first hint is provided by the study of the energy balance of the
system \cite{Lucha:2021mwx}. This is most easily done in the
\textit{static limit} of
the theory, with very heavy quarks, fixed at spatial positions.
The system would choose configurations with \textit{minimal energy}.
\par
The problem is analytically solved in the strong coupling limit of
lattice theory \cite{Dosch:1982ep} and confirmed by direct lattice
numerical calculations \cite{Alexandrou:2004ak,Okiharu:2004wy,
Okiharu:2004ve,Suganuma:2011ci,Cardoso:2011fq,Bicudo:2017usw}.
In the strong coupling limit, the potential energy is concentrated
on the \textit{Wilson lines}, which are the path-ordered gluon field
phase factors, with constant linear energy density.
\par
Figure \rf{f1} represents schematic geometric configurations of
the two types of structure. In Fig. \rf{f1}a, one has a
compact tetraquark, formed by confining interactions, through
diquark--antidiquark global interaction. In Fig. \rf{f1}b, one
has two meson clusters.
\bfg 
%\vspace*{1 cm}
\bc
\includegraphics[scale=0.7]{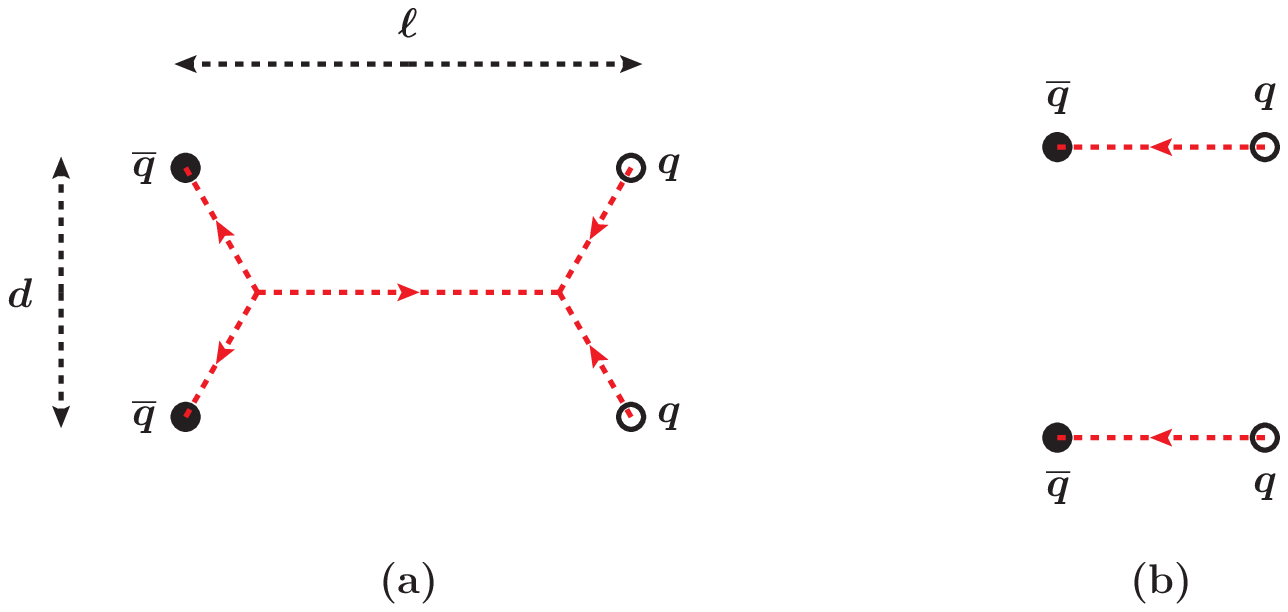}
\ec
\caption{(a): Compact tetraquark, formed by confining interactions,
through diquark--antidiquark global interaction.
(b): Two meson clusters.}
\lb{f1}
\efg
\par
The compact tetraquark formation is energetically favored if the
diquark interdistance $d$ is much smaller than the
quark-antiquark interdistance $\ell$:
\be \lb{e4}
d\ll \ell.
\ee
\par
This shows that both structures have nonzero probabilities to be
formed. However, quarks have finite masses and move in space. Even
if a compact tetraquark has been formed, there is a sizable
probability that the quarks, during their motion, reach the
two-meson-cluster configuration, in which case the system dislocates
or decays.
\par
An interesting and nontrivial case is when the compact
tetraquark mass lies below the two-meson threshold, which prevents
dislocation. Nevertheless, there will be, through fluctuations,
constant transitions to the two-meson virtual states. This
will have the tendancy to expand the compact system into a more
loosely bound system, close to a molecular-type state.
This is a dynamical mechanism, whose precise outcome necessitates
the resolution of the four-body bound state problem, in the presence
of the confining forces. For the time being, this problem has not yet
been satisfactorily solved. One is obliged, to analyze the problem
with sufficient accuracy, to have recourse to approximation schemes.
We shall consider below the framework provided by the effective
field theory approach. A detailed account of this work can be found
in \cite{Sazdjian:2022kaf}.
\par

\section{Effective field theory approach} \lb{s3}

According to the energy balance analysis, there is
always a probability that a compact tetraquark be formed from
the action of the confining forces (actually an infinite tower
of such states). We assume that in first approximation, because of
the compactness of the state, the latter can be assimilated to a
pointlike object and treated as an elementary particle.
\par
We are mainly interested in the case where the mass of the latter
particle lies below the lowest possible two-meson threshold. This
might be the case when the compact tetraquark is the ground state
of the corresponding spectrum.
\par
The two mesons are designated by $M_1^{}$ and $M_2^{}$,
with masses $m_1^{}$ and $m_2^{}$, respectively.
The bare mass of the tetraquark is $m_{t0}^{}$.
\par
Quark flavors and spins will be ignored, as not playing a
fundamental role here.
\par
Because of the initial general structure of the tetraquark,
containing two meson clusters, the pointlike tetraquark has
necessarily a coupling to the mesons $M_1^{}$ and
$M_2^{}$. The coupling is assumed scalar
and is designated by $g'$; the latter is dimensionless and
is factored by the mass term $(m_1^{}+m_2^{})$.
\par
The coupling $g'$ generates, through meson loops, radiative
corrections inside the tetraquark propagator,
modifying the parameters of the bare propagator. In particular, the
bare mass $m_{t0}^{}$ gets  changed into a physical
mass $m_{t}^{}$.
This is graphically represented in Fig. \rf{f2}.
\bfg 
%\vspace*{1 cm}
\bc
\includegraphics[scale=0.8]{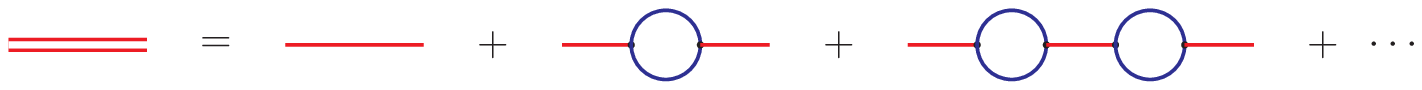}
\ec
\caption{The full tetraquark propagator, including the radiative
corrections, coming from the coupling to the two meson clusters.}
\lb{f2}
\efg
\par
Mutual interactions of mesons are neglected (in first approximation);
they play a less important role than the direct tetraquark-two-meson
coupling.
\par
The full tetraquark propagator becomes
\be \lb{e5}
D_{t}^{}(s)=\frac{i}{s-m_{t0}^2+i(m_1^{}+m_2^{})^2g^{\prime 2}J(s)},
\ee
where $s$ stands for $p^2$ and $J$ is the standard loop function.
The divergence of $J$ is absorbed by the bare mass
term, yielding a renormalized mass $m_{t1}^{}$:
\be \lb{e6}
m_{t1}^2=m_{t0}^2-i (m_1^{}+m_2^{})^2g^{\prime 2}J^{\mathrm{div}}.
\ee
\par
The renormalized tetraquark propagator is now
\be \lb{e7}
D_{t}^{}(s)=\frac{i}{s-m_{t1}^2+i(m_1^{}+m_2^{})^2g^{\prime 2}
J^{\mathrm{r}}(s)},
\ee
where $J^{\mathrm{r}}$ is the renormalized finite part of $J$.
Notice that $g'$ does not undergo any renormalization.
The mass term $m_{t1}^{}$ does not yet represent the
physical mass of the tetraquark. The latter is determined from the
pole position of the propagator.
\par
We stick here to the case of heavy quarks and heavy mesons, treating
the problem in its nonrelativistic limit, referred to the
two-meson threshold. The nonrelativistic energy $E$
is introduced through the standard definition
\be \lb{e8} 
\sqrt{s}=(m_1^{}+m_2^{})+E.
\ee
\par
The nonrelativistic energy corresponding to the renormalized mass
$m_{t1}^{}$ of the tetraquark is defined similarly:
\be \lb{e9}
E_{t1}^{}=m_{t1}-(m_1^{}+m_2^{}),\ \ \ \ \ E_{t1}^{}<0.
\ee
Since we are considering the case of stable tetraquarks (under the
strong interactions), the mass of the tetraquark is expected to lie
below the two-meson threshold.
The physical nonrelativistic energy of the tetraquark will be
designated by $E_{t}^{}$ and is also expected to be negative.
\par
To simplify notations, we shall use henceforth the reduced
dimensionless energy variables $e$ through the definitions 
\be \lb{e10} 
e\equiv\frac{E}{2m_r^{}},\ \ \ \ 
e_{t1}^{}\equiv\frac{E_{t1}^{}}{2m_r^{}},\ \ \ \ 
e_{t}^{}\equiv\frac{E_{t}^{}}{2m_r^{}},\ \ \ \
m_r^{}=\frac{m_1^{}m_2^{}}{(m_1^{}+m_2^{})}.
\ee
\par
The quantity $-e_{t}^{}$ represents the nonrelativistic
binding energy of the tetraquark by reference to the two-meson
threshold, although it is different from the binding energy
defined from the confining forces by reference to the quark masses.
\par
One finds for the nonrelativistic energy $e_{t}^{}$
of the tetraquark the equation
\be \lb{e11} 
-e_{t}^{}+e_{t1}^{}+\frac{g^{\prime 2}}{16\pi}\sqrt{-e_{t}^{}}=0,
\ee
whose solution is
\be \lb{e12} 
\sqrt{-e_{t}^{}}=\frac{1}{2}\Big[-\frac{g^{\prime 2}}{16\pi}+
\sqrt{\Big(\frac{g^{\prime 2}}{16\pi}\Big)^2-4e_{t1}^{}}\Big].
\ee
\bfg 
%\vspace*{1 cm}
\bc
\includegraphics[scale=0.65]{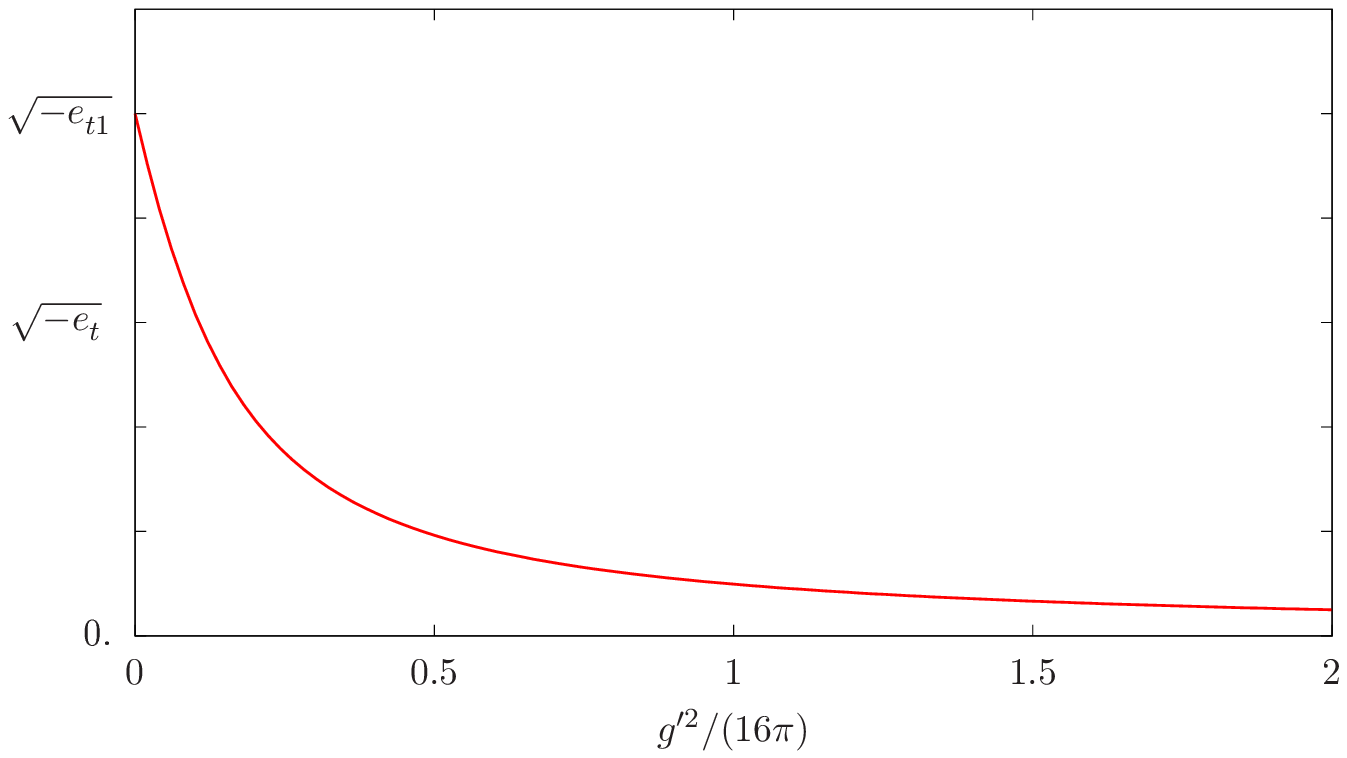}
\ec
\caption{Variation of the square-root of the binding energy as a
function of $g^{\prime 2}/(16\pi)$.}
%The value of
%$-e_{tc1}$ ($\equiv -E_{tc1}^{}/(2m_r^{})$ has been fixed at 0.01.}  
\lb{f3}
\efg
The binding energy $-e_{t}^{}$ is a decreasing function of
$g^{\prime 2}/(16\pi)$ and comes out smaller than
$-e_{t1}^{}$, reaching the value 0 when $g'\rightarrow\infty$.
This is represented in Fig. \rf{f3}.
\par
We find a very rapid decrease of the binding energy. With ordinary
values of $g^{\prime 2}/(16\pi)$, of the order of 1, the binding
energy decreases by a factor of 1/100. The state takes the appearance
of a \textit{shallow} bound state.
\par

\section{Compositeness} \lb{s4}

The comparison of the molecular and compact schemes is reminiscent
of a general problem, already raised in the past in the case of the
deuteron state, denoted under the term of \textit{compositeness}
\cite{Weinberg:1965zz}.
\par 
Weinberg has shown that this question can receive, in the
nonrelativistic limit, a precise and model-independent answer, by
relating the probability of a state as being elementary
(or compact) to observable quantities, represented by the scattering
length and the effective range of the two constituents of the
molecular scheme in the $S$-wave state of their scattering amplitude. 
Designating by $Z$ this probability, one has the following relations
for the scattering length $a$ and the effective range $r_e^{}$,
adapted to the tetraquark problem:
\be \lb{e13} 
a=\frac{2(1-Z)}{(2-Z)}R,\ \ \ \ \ 
r_e^{}=-\frac{Z}{(1-Z)}R,\ \ \ \ \ 
R=(-2m_r^{}E_t^{})^{-1/2},
\ee
where $R$ is the radius of the bound state, $m_r^{}$ the reduced
mass of the two-meson system and $E_t^{}$ the tetraquark
nonrelativistic energy.
\par
Investigations about the compositeness criterion and its
applicability to various tetraquark candidates, as well as to
ordinary hadrons, can be found in
\cite{Baru:2003qq,Cleven:2011gp,Hanhart:2011jz,
Hyodo:2011qc,Aceti:2012dd,Sekihara:2014kya,Guo:2015daa,Kang:2016ezb,
Meissner:2015mza,Oller:2017alp,Guo:2020vmu,Esposito:2021vhu,
Li:2021cue,Baru:2021ldu,Kinugawa:2021ykv,Song:2022yvz}.
\par
The contribution of the tetraquark state, in the $s$-channel, to
the two-meson elastic scattering amplitude is obtained by inserting
the tetraquark propagator between two tetraquark-two-meson couplings
(cf. Fig. \rf{f4}).
\bfg
%\vspace*{0.25 cm}
\bc
\includegraphics[scale=0.9]{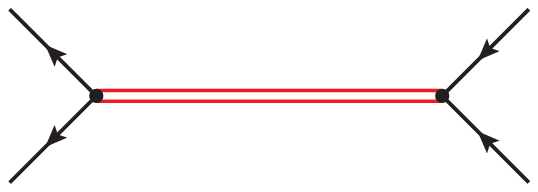}
%\vspace*{0.25 cm}
\ec
\caption{The tetraquark contribution, in the $s$-channel, to the
two-meson elastic scattering amplitude.}
\lb{f4}
\efg
\par
From the scattering length and the effective range one obtains
$Z$:
\be \lb{e14}
Z=\frac{\sqrt{-e_{t}^{}}}{\sqrt{-e_{t}^{}}+\frac{1}{2}
\frac{g^{\prime 2}}{16\pi}}.
\ee
When $Z=1$, we have the case of a pure compact tetraquark, while
for $Z=0$, we have a pure molecular state.      
The variation of $Z$ with respect to $g^{\prime 2}/(16\pi)$ is
represented in Fig. \rf{f5}.
%\par
%\newpage
\bfg
%\vspace*{0.25 cm}
\bc
\includegraphics[scale=0.65]{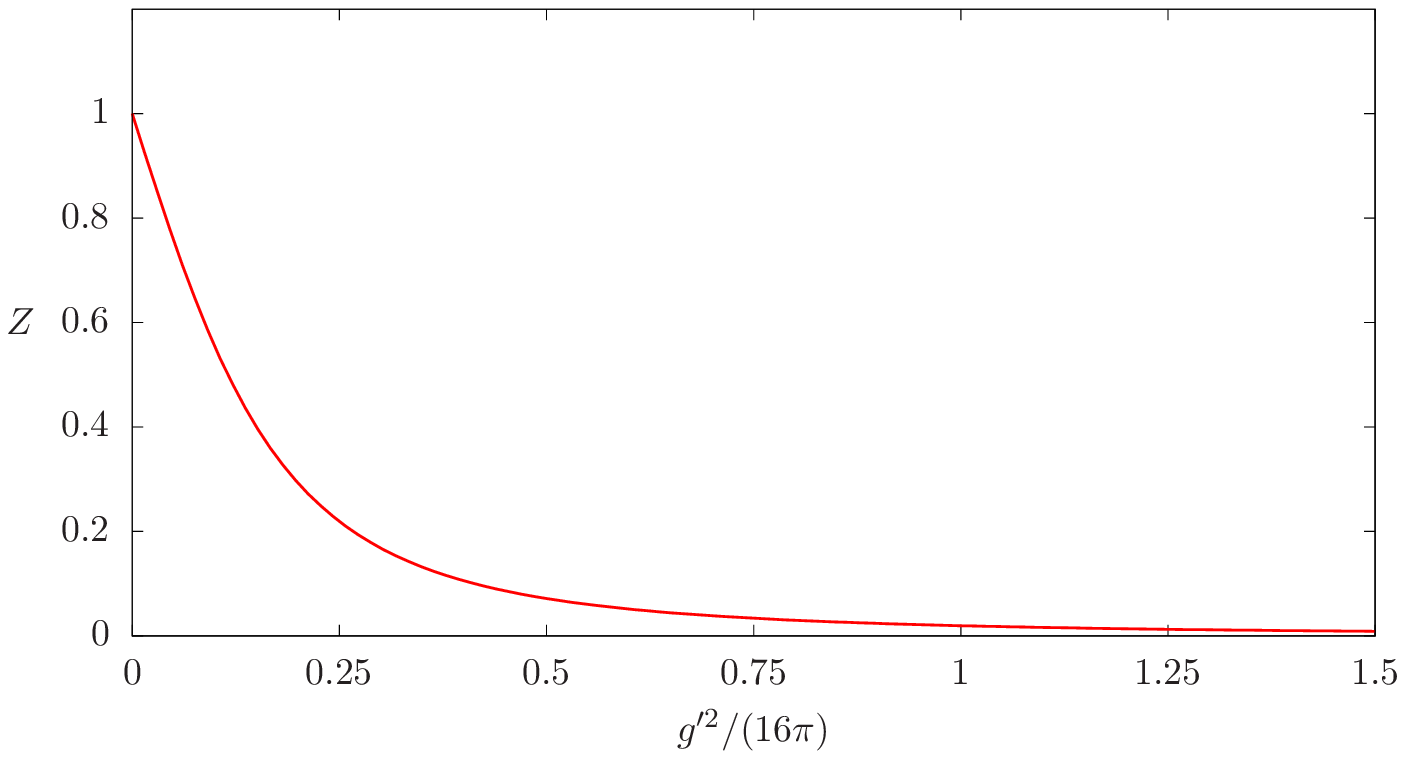}
%\vspace*{0.25 cm}
\ec
\caption{Variation of $Z$ as a function of $g^{\prime 2}/(16\pi)$.}
%$-e_{tc1}^{}$ has been fixed at 0.01.
\lb{f5}
\efg
\par
Like the binding energy, $Z$ is a rapidly decreasing function of
$g^{\prime 2}/(16\pi)$. For values of the latter of the order of 1,
$Z$ is very close to zero. This confirms the interpretation that
the coupling of the compact tetraquark to its internal mesonic
clusters qualitatively deforms its initial structure, bringing it
into a form closer to a molecular-type state.
\par

\section{Conclusion} \lb{s5}

A compact tetraquark, formed from the confining forces acting
between quarks and gluons, rapidly evolves, under the influence of
the clustering phenomenon, towards a molecular-type state. The
origin of the state is, however, of compact nature: nowhere, in
the present model, did we consider direct interactions between mesons.
\par
The experimental test for this phenomenon is provided by the
measure of the elementariness coefficient $Z$.
Pure molecular states are characterized by the value
$Z=0$. A value $Z\neq 0$, beyond uncertainties, reflects
the existence of an original compact tetraquark. Many tetraquark
candidates fall in this category.
\par
Because of the shrinking of the tetraquark binding energy to values
close to zero, shallowness of many bound states may receive a natural
explanation from the above mechanism.
\par
The present study can also be extended to the case of resonances and
to the case when direct meson-meson interactions are incorporated.
\par

\vspace{0.25 cm}
\noindent
\textbf{Funding}. This research has received financial
support from the EU research and innovation programme Horizon 2020,
under Grant agree\-ment No. 824093, and from the joint CNRS/RFBR
Grant No. PRC Russia/19-52-15022.
\par

\end{document}